\documentclass[letterpaper,10pt]{article} 

\usepackage{opticameet3} 

\newcommand\authormark[1]{\textsuperscript{#1}}

\usepackage{amsmath,amssymb}
\usepackage{xspace}
\usepackage{xcolor}
\usepackage{tikz}
\usepackage{wrapfig}
\usetikzlibrary{arrows.meta,calc,backgrounds}
\definecolor{planezero}{HTML}{0072B2}
\definecolor{planeone}{HTML}{D55E00}
\usepackage{etoolbox}
\makeatletter
\long\def\@makecaption#1#2{%
  \footnotesize
  \vskip\abovecaptionskip
  \sbox\@tempboxa{#1. #2}%
  \ifdim \wd\@tempboxa >\linewidth
    \parbox{\linewidth}{#1. #2}%
  \else
    \global \@minipagefalse
    \hb@xt@\hsize{\hfil\box\@tempboxa\hfil}%
  \fi
  \vskip\belowcaptionskip\normalsize}
\makeatother
\usepackage[colorlinks=true,bookmarks=false,citecolor=blue,urlcolor=blue]{hyperref} 

\newcommand{\sysname}{OMRC\xspace}

\providecommand{\alltoall}{\textsc{AllToAll}\xspace}
\providecommand{\allreduce}{\textsc{AllReduce}\xspace}
\providecommand{\allgather}{\textsc{AllGather}\xspace}

\providecommand{\reducescatter}{\textsc{ReduceScatter}\xspace}
\providecommand{\myparab}[1]{\vspace{0pt}\noindent\textbf{#1} }

\providecommand{\eg}{\emph{e.g.,} }

\begin{document}

\title{OMRC: Multipath Reliability for Direct-Connect Optical Interconnects}

\vspace{-15pt} 
\author{Eric Ding,\authormark{1,*} Keshav Pradeep,\authormark{1} and Rachee Singh\authormark{1}}

\address{\authormark{1}Department of Computer Science, Cornell University, 107 Hoy Rd, Ithaca, NY 14853, USA}

\email{\authormark{*}ericding@cs.cornell.edu} 


\vspace{-15pt} 
\begin{abstract}
Network failures degrade Multipath Reliable Connection (MRC) in direct-connect optical datacenters. We propose Optical MRC (OMRC), which remaps collectives and circuits, preserving 95.7\% of training throughput, restoring stalled inference, and achieving 4.7$\times$ MRC’s simulated bandwidth.

\end{abstract}
\vspace{-6pt} 

\section{Introduction}
\def\mrcFourHops{2}
\def\omrcFourHops{1}
\def\mrcSevenHops{3}
\def\omrcSevenHops{1}

Distributed training and inference jobs exchange data among hundreds of accelerators through collective communication operations like \allreduce and \alltoall. Even a single failed network link between two accelerators stalls the collective and the training or inference step that waits on the collective. Unfortunately, link failures occur frequently in production AI clusters due to link flaps, network interface card (NIC) and switch faults, and laser failures~\cite{mrc}. While the ML application can recover from such failures by restarting its processes on a new set of accelerators and rebuilding inter-accelerator connections, such a restart is expensive at scale.

\begin{figure}[h!]
  \centering
  \setlength{\abovecaptionskip}{3pt}\setlength{\belowcaptionskip}{0pt}
  \definecolor{tpgray}{gray}{0.55}
  \definecolor{failred}{HTML}{B2182B}
  \providecommand{\failx}[1]{%
    \begin{scope}[shift={(#1)}]
      \fill[white] (0,0) circle (2.8pt);
      \draw[failred,line width=1.2pt,line cap=round]
        (-2.1pt,-2.1pt)--(2.1pt,2.1pt) (-2.1pt,2.1pt)--(2.1pt,-2.1pt);
    \end{scope}}
  \resizebox{\linewidth}{!}{%
\begin{tikzpicture}[
  x=1cm,y=1cm,font=\scriptsize,
  gpu/.style={circle,draw=black!70,fill=white,minimum size=0.44cm,inner sep=0pt},
  sw/.style={draw=#1,line width=0.9pt,fill=white,rounded corners=1.5pt,minimum width=1.12cm,minimum height=0.38cm,inner sep=0pt},
  p0/.style={planezero,line width=0.9pt},
  p1/.style={planeone,line width=0.9pt},
  faint/.style={opacity=0.22},
  send/.style={line width=1.8pt,-{Stealth[length=2.1mm,width=2.1mm]}},
  tp/.style={tpgray,line width=0.75pt},
  dp/.style={planezero,line width=0.9pt},
  pcap/.style={text width=4.1cm,align=center,anchor=north,inner sep=0pt,execute at begin node={\hyphenpenalty=10000\exhyphenpenalty=10000}},
  head/.style={font=\small\bfseries,anchor=base},
  rowlab/.style={rotate=90,align=center,font=\scriptsize},
]
\def\s{1.1}
\def\cZ{1.75} \def\cA{6.6} \def\cB{10.85} \def\cC{15.1}
\def\rA{2.35} \def\rB{0}
\def\hy{\rA+\s+0.36}

\node[head] at (\cZ,\hy) {Electrical vs.\ optical DC};
\node[head] at (\cA,\hy) {Topology and failure};
\node[head] at (\cB,\hy) {MRC};
\node[head] at (\cC,\hy) {OMRC (this work)};

\draw[black!35,dashed,line width=0.5pt] (3.72,\rB-0.95) -- (3.72,\rA+\s+0.2);
\node[rowlab] at (4.2,\rA+0.25) {\textbf{Limitation 1}\\different ring orders};
\node[rowlab] at (4.2,\rB+0.25) {\textbf{Limitation 2}\\switch failure};

\foreach \base/\tag/\lab in {\rA/x/Clos,\rB/o/OCS}{
  \foreach \i in {0,1,2,3}{\node[gpu] (\tag\i) at (\cZ-1.2+0.8*\i,\base) {\i};}
  \node[sw=planezero] (\tag s0) at (\cZ-0.62,\base+0.92) {\lab\ 0};
  \node[sw=planeone]  (\tag s1) at (\cZ+0.62,\base+0.92) {\lab\ 1};
  \foreach \i in {0,1,2,3}{
    \draw[p0] (\tag\i.north) -- ($(\tag s0.south west)!{(\i+0.5)/4}!(\tag s0.south east)$);
    \draw[p1] (\tag\i.north) -- ($(\tag s1.south west)!{(\i+0.5)/4}!(\tag s1.south east)$);
  }
}
\failx{$(x0.north)!0.45!($(xs0.south west)!0.125!(xs0.south east)$)$}
\draw[p1,send] (x0.north) -- ($(xs1.south west)!0.125!(xs1.south east)$);
\draw[p1,send] ($(xs1.south west)!0.375!(xs1.south east)$) -- (x1.north);
\node[pcap,text width=3.55cm] at (\cZ,\rA-0.27) {\textbf{Electrical DC.} Ports connect to packet-switched planes.};
\node[pcap,text width=3.55cm] at (\cZ,\rB-0.27) {\textbf{Direct-connect optical DC.} Ports connect to OCSes. Each OCS forms one ring, called a plane.};

\foreach \cx/\tag in {\cA/a,\cB/b,\cC/c}{
  \begin{scope}[shift={(\cx-0.6,\rA)}]
    \node[gpu] (\tag0) at (0,\s) {0};
    \node[gpu] (\tag1) at (\s,\s) {1};
    \node[gpu] (\tag2) at (\s,0) {2};
    \node[gpu] (\tag3) at (0,0) {3};
  \end{scope}}

\draw[p0] (a0)--(a1); \draw[p0] (a1)--(a2); \draw[p0] (a2)--(a3); \draw[p0] (a3)--(a0);
\draw[p1] (a0)--(a2); \draw[p1] (a2) to[bend right=40] (a1);
\draw[p1] (a1)--(a3); \draw[p1] (a3) to[bend left=40] (a0);
\failx{$(a0)!0.5!(a1)$}
\node[pcap] at (\cA,\rA-0.27) {Planes 0 and 1 order the GPUs 0,1,2,3 and 0,2,1,3. Link 0--1 fails in plane 0.};

\draw[p0] (b0)--(b1); \draw[p0] (b1)--(b2); \draw[p0] (b2)--(b3); \draw[p0] (b3)--(b0);
\draw[p1,faint] (b1)--(b3); \draw[p1,faint] (b3) to[bend left=40] (b0);
\draw[p1,send] (b0)--(b2);
\draw[p1,send] (b2) to[bend right=40] (b1);
\failx{$(b0)!0.5!(b1)$}
\node[planeone,font=\scriptsize\bfseries,anchor=west] at ($(b1)!0.5!(b2)+(0.38,0)$) {2 hops};
\node[pcap] at (\cB,\rA-0.27) {Since the ring keeps order 0,1,2,3, the send $0\to1$ detours through GPU~2.};

\draw[p0,faint] (c0)--(c1); \draw[p0,faint] (c1)--(c2); \draw[p0,faint] (c2)--(c3); \draw[p0,faint] (c3)--(c0);
\draw[p1,send] (c0)--(c2);
\draw[p1,send] (c2) to[bend right=40] (c1);
\draw[p1,send] (c1)--(c3);
\draw[p1,send] (c3) to[bend left=40] (c0);
\failx{$(c0)!0.5!(c1)$}
\node[planeone,font=\scriptsize\bfseries,anchor=west] at ($(c1)!0.5!(c2)+(0.38,0)$) {1 hop};
\node[pcap,text width=4.5cm] at (\cC+0.1,\rA-0.27) {OMRC moves the channel to plane 1 with order 0,2,1,3. Each send takes one hop.};

\foreach \cx/\tag in {\cA/d,\cB/e,\cC/f}{
  \begin{scope}[shift={(\cx-0.6,\rB)}]
    \node[gpu] (\tag0) at (0,\s) {0};
    \node[gpu] (\tag1) at (\s,\s) {1};
    \node[gpu] (\tag2) at (0,0) {2};
    \node[gpu] (\tag3) at (\s,0) {3};
  \end{scope}}

\newcommand{\tplinks}[3]{
  \foreach \dy in {#3}{\draw[tp] ($(#1.east)+(0,\dy)$)--($(#2.west)+(0,\dy)$);}}

\tplinks{d0}{d1}{0.085,0,-0.085} \tplinks{d2}{d3}{0.085,0,-0.085}
\draw[dp] (d0)--(d2); \draw[dp] (d1)--(d3);
\failx{$(d0)!0.5!(d2)$}
\failx{$(d1)!0.5!(d3)$}
\node[pcap] at (\cA,\rB-0.27) {TP: three planes per pair; DP: one. A switch failure breaks both 0--2 and 1--3.};

\tplinks{e0}{e1}{0.085,0,-0.085} \tplinks{e2}{e3}{0.085,0,-0.085}
\draw[dp] (e0)--(e2); \draw[dp] (e1)--(e3);
\failx{$(e0)!0.5!(e2)$}
\failx{$(e1)!0.5!(e3)$}
\node[failred,font=\scriptsize\bfseries,anchor=east] at ($(e0)!0.5!(e2)+(-0.2,0)$) {stall};
\node[failred,font=\scriptsize\bfseries,anchor=west] at ($(e1)!0.5!(e3)+(0.2,0)$) {stall};
\node[pcap] at (\cB,\rB-0.27) {Both DP pairs lose their only plane. MRC cannot use the TP planes.};

\tplinks{f0}{f1}{0.045,-0.045} \tplinks{f2}{f3}{0.045,-0.045}
\draw[dp] (f0)--(f2); \draw[dp] (f1)--(f3);
\failx{$(f0)!0.5!(f2)$}
\failx{$(f1)!0.5!(f3)$}
\draw[planeone,line width=1.8pt] ($(f0.south west)+(-0.02,0)$) to[bend right=22] ($(f2.north west)+(-0.02,0)$);
\draw[planeone,line width=1.8pt] ($(f1.south east)+(0.02,0)$) to[bend left=22] ($(f3.north east)+(0.02,0)$);
\node[planeone,font=\scriptsize\bfseries,anchor=center,align=center] at ($(f0)!0.5!(f3)$) {moved\\planes};
\node[pcap] at (\cC,\rB-0.27) {The OCS reallocates one plane from each TP pair to restore both DP pairs.};

\node[font=\small,anchor=base] at (\cZ,\rB-1.62) {(a)};
\node[font=\small,anchor=base] at (\cB,\rB-1.62) {(b)};

\begin{scope}[shift={(4.0,\rB-1.12)}]
  \failx{0.15,0} \node[anchor=west] at (0.3,0) {failed link};
  \draw[p0] (1.95,0)--(2.35,0); \node[anchor=west] at (2.4,0) {plane 0 / DP plane};
  \draw[p1] (5.05,0)--(5.45,0); \node[anchor=west] at (5.5,0) {plane 1 / moved plane};
  \draw[tp] (8.7,0)--(9.1,0); \node[anchor=west] at (9.15,0) {TP plane};
  \draw[send,planeone] (10.7,0)--(11.2,0); \node[anchor=west] at (11.25,0) {Transfer};
\end{scope}
\end{tikzpicture}}%
  \caption{(a) Electrical and direct-connect optical datacenters (DCs). NIC ports connect to packet-switched planes in an electrical DC and to optical circuit switches (OCSes) in a direct-connect optical DC. (b) Link and switch failures in a direct-connect optical DC. Each row shows a limitation of Multipath Reliable Connection (MRC) on direct-connect optical DCs and how \sysname addresses it. A channel is a slice of a collective that follows its own ring order. TP and DP denote tensor-parallel and data-parallel groups.}
  \label{fig:overview}
\end{figure}
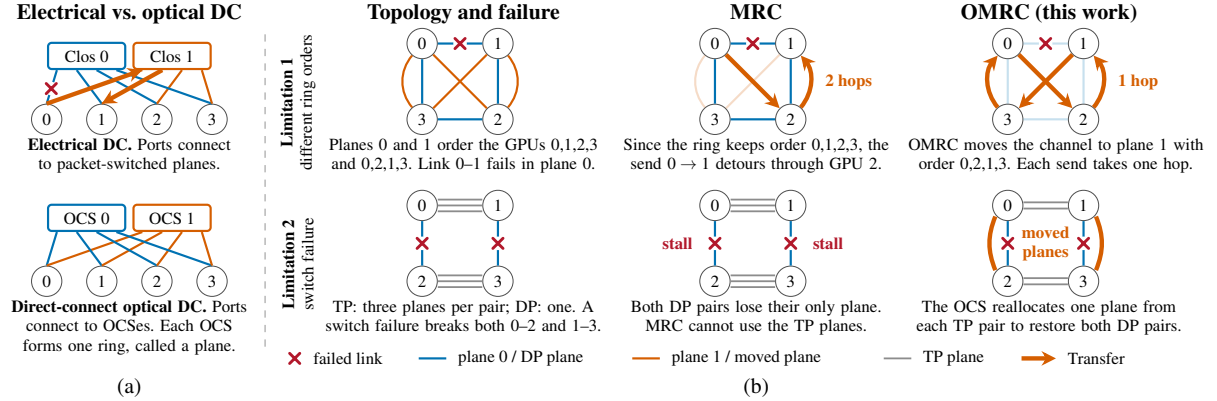

Hyperscalers avoid job restarts by handling link failures in the network transport with Multipath Reliable Connection (MRC)~\cite{mrc}. MRC runs on a multi-plane network in which the NIC of each accelerator splits its bandwidth across several lower-speed ports (Fig.~\ref{fig:overview}(a), top). For example, an 800~Gb/s NIC operates as eight 100~Gb/s ports~\cite{mrc}, each connected to a different plane. A plane is an independent packet-switched network, typically a two-tier Clos, that connects one port of each NIC in the cluster. Since all planes have the same structure, a path between two accelerators has the same length in every plane. MRC sprays the packets of a connection between two accelerators across the planes. When one path fails, MRC stops sending on that path and moves its traffic to the remaining paths. By design, the remaining paths have the same length as the failed path, so the connection loses only the failed path's bandwidth.

Since electrical packet switches in large ML datacenters (DCs) consume massive amounts of energy, recent work has proposed replacing them with optical circuit switches (OCSes) to achieve orders-of-magnitude efficiency gains~\cite{opus}. In the resulting direct-connect datacenters, OCSes connect accelerator NICs with optical circuits, without any packet switches in between (Fig.~\ref{fig:overview}(a), bottom). To maximize collective performance, network operators configure each OCS to connect two NIC ports of each accelerator into a ring, called a plane~\cite{topoopt,opus}. Therefore, an accelerator with degree $D$ ($D$ NIC ports) belongs to $D/2$ planes. Ring collectives like \allreduce run on all planes in parallel. Each step of a ring collective sends data to a neighbor in one hop within individual planes. 
Operators order the accelerators differently in different planes. With different orders, each accelerator has more neighbors and shorter paths to the other accelerators, which benefits traffic like \alltoall. In Fig.~\ref{fig:overview}(b), each GPU has degree four and belongs to two planes (plane~0 orders the GPUs 0,1,2,3 and plane~1 orders them 0,2,1,3). Each communication group, the set of accelerators that communicate for one type of parallelism, has a dedicated subset of the planes~\cite{actina}. For example, data-parallel (DP) and tensor-parallel (TP) groups use separate planes. 

In this work, we show that, when applied to direct-connect optical datacenters, MRC reroutes around failed links with two important limitations. First, the planes of a direct-connect DC do not offer equal-length paths between a pair of GPUs. Two GPUs that are neighbors in one plane can be several hops apart in another plane. If link 0--1 fails in plane~0, MRC forwards its traffic through GPU~2 in plane~1 (Fig.~\ref{fig:overview}(b), top row). This detour consumes the bandwidth of an extra link and adds latency. Second, a group that loses all of its planes stalls with MRC (\eg due to switch failures), even when other groups have healthy planes (Fig.~\ref{fig:overview}(b), bottom row).

We propose Optical Multipath Reliable Connection (\sysname), a control layer that extends the MRC transport for direct-connect optical DCs. When a link fails, \sysname remaps collectives onto intact planes (Fig.~\ref{fig:overview}(b), top right). If a group loses connectivity, \sysname moves planes between communication groups by reconfiguring the topology (Fig.~\ref{fig:overview}(b), bottom right). We implemented \sysname in MSCCL++ and evaluated it with training and inference jobs of Qwen1.5-MoE, a 14.3B-parameter mixture-of-experts (MoE) model, on four NVIDIA L40 GPUs connected through a Polatis OCS. When one link fails for 40~s during training, \sysname remaps the affected channel and improves training throughput by 5.3\% over MRC. When the job loses its sole TP plane during inference, MRC stalls. \sysname moves one plane from the key-value (KV) cache transfer group to restore TP group connectivity, retaining 99.3\% of pre-failure inference throughput. Large-scale simulations show that \sysname achieves up to $4.74\times$ MRC's algorithmic bandwidth in networks with multiple failures.


\section{Design of \sysname}
\sysname is a topology-aware control layer that extends MRC for AI workloads on direct-connect OCS networks. Its core contribution is to adapt collective schedules to the physical topology after failures, reducing intermediate-node forwarding, and to adapt the physical topology when the affected communication group lacks connectivity.

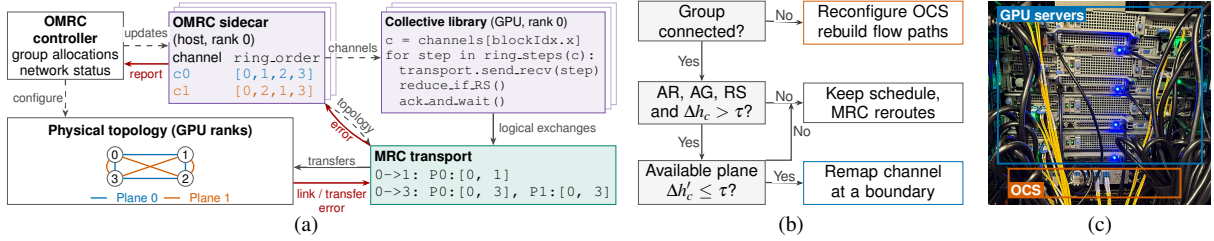
\begin{figure}[h!]
  \centering
  \definecolor{hostgpu}{HTML}{7952A3}
  \definecolor{nictransport}{HTML}{00856A}
  \resizebox{\linewidth}{!}{%
  \begin{tikzpicture}[
    x=1cm,y=.8cm,font=\sffamily\fontsize{8}{8.6}\selectfont,
    box/.style={draw=black!65,fill=white,align=center,inner sep=2pt},
    compute/.style={box,draw=hostgpu,fill=hostgpu!10},
    network/.style={box,draw=nictransport,fill=nictransport!12},
    ctl/.style={black!70,line width=.65pt,dashed,-{Stealth[length=1.5mm]}},
    payload/.style={black!70,line width=.75pt,-{Stealth[length=1.5mm]}},
    fault/.style={red!65!black,line width=.8pt,-{Stealth[length=1.5mm]}},
    edgecaption/.style={font=\sffamily\fontsize{7}{7.6}\selectfont,fill=white,inner sep=.7pt},
    decision/.style={box,fill=black!4,text width=2.2cm,minimum height=.55cm,
                     font=\sffamily\fontsize{9.5}{10.3}\selectfont,inner sep=2pt},
    action/.style={box,text width=2.8cm,minimum height=.55cm,
                   font=\sffamily\fontsize{9.5}{10.3}\selectfont,inner sep=2pt},
    branch/.style={black!75,line width=.65pt,-{Stealth[length=1.3mm]}}
  ]
    \begin{scope}[xshift=0cm]
    \node[network,minimum width=3.9cm,minimum height=.85cm,align=left,
          font=\ttfamily\fontsize{8}{8.6}\selectfont] (transport) at (8.95,.2) {%
      \begin{tabular}{@{}l@{}}
        {\sffamily\bfseries MRC transport}\\[1pt]
        0->1: P0:[0, 1]\\
        0->3: P0:[0, 3], P1:[0, 3]
      \end{tabular}};
    \node[compute,minimum width=3.9cm,align=left,
          font=\ttfamily\fontsize{7.2}{8}\selectfont] (library) at (8.95,2.8) {%
      \begin{tabular}{@{}l@{}}
        {\sffamily\textbf{Collective library} (GPU, rank 0)}\\[1pt]
        c = channels[blockIdx.x]\\
        for step in ring\_steps(c):\\
        \quad transport.send\_recv(step)\\
        \quad reduce\_if\_RS()\\
        \quad ack\_and\_wait()
      \end{tabular}};

    \node[compute,minimum width=2.8cm,align=left,anchor=north,
          font=\ttfamily\fontsize{8}{8.6}\selectfont] (sidecar)
      at (4.4,0 |- library.north) {%
      \setlength{\tabcolsep}{3pt}%
      \begin{tabular}{@{}ll@{}}
        \multicolumn{2}{@{}l@{}}{{\sffamily\bfseries OMRC sidecar}}\\
        \multicolumn{2}{@{}l@{}}{{\sffamily (host, rank 0)}}\\[1pt]
        {\sffamily channel} & ring\_order\\
        \textcolor{planezero}{c0} & \textcolor{planezero}{[0,1,2,3]}\\
        \textcolor{planeone}{c1} & \textcolor{planeone}{[0,2,1,3]}
      \end{tabular}};

    \node[box,text width=2.0cm,anchor=north] (controller)
      at (1.05,0 |- library.north)
      {\textbf{OMRC}\\\textbf{controller}\\group allocations\\network status};

    \begin{scope}[on background layer]
      \foreach \offset in {.20,.10} {
        \draw[box,draw=hostgpu!50]
          ($(sidecar.south west)+(\offset,\offset)$)
          rectangle ($(sidecar.north east)+(\offset,\offset)$);
        \draw[box,draw=hostgpu!50]
          ($(library.south west)+(\offset,\offset)$)
          rectangle ($(library.north east)+(\offset,\offset)$);
      }
    \end{scope}
    \draw[ctl] (controller.east)--node[edgecaption,above,yshift=2pt] {updates}
      (sidecar.west |- controller.east);
    \coordinate (report-in) at ($(controller.east)+(0,-.42)$);
    \draw[fault] (sidecar.west |- report-in)
      --node[edgecaption,below,yshift=-2pt,text=red!65!black] {report} (report-in);
    \draw[ctl] (sidecar.east)--node[edgecaption,above,yshift=2pt] {channels}
      (library.west |- sidecar.east);
    \draw[ctl] (sidecar.south east) to[bend left=12]
      node[edgecaption,sloped,above] {topology} (transport.north west);
    \draw[fault] (transport.north west) to[bend left=12]
      node[edgecaption,sloped,below,text=red!65!black] {error} (sidecar.south east);

    \path let \p1=(controller.west), \p2=(sidecar.east),
              \n1={\x2-\x1-.6cm} in
      node[box,minimum width=\n1,minimum height=1.6cm,anchor=south west]
        (topology) at (controller.west |- transport.south) {};
    \begin{scope}[shift={(topology.center)},
      rank/.style={circle,draw=black!65,fill=white,inner sep=0pt,
                   minimum size=.29cm,font=\sffamily\fontsize{7}{7.6}\selectfont},
      ring/.style={line width=.7pt}]
      \node[font=\sffamily\bfseries\fontsize{8}{8.6}\selectfont]
        at (0,.68) {Physical topology (GPU ranks)};
      \node[rank] (gpu0) at (-.65,.15) {0};
      \node[rank] (gpu1) at (.65,.15) {1};
      \node[rank] (gpu2) at (.65,-.4) {2};
      \node[rank] (gpu3) at (-.65,-.4) {3};
      \foreach \a/\b in {0/1,1/2,2/3,3/0}
        \draw[ring,planezero] (gpu\a)--(gpu\b);
      \draw[ring,planeone] (gpu0)--(gpu2);
      \draw[ring,planeone] (gpu2) to[bend right=45] (gpu1);
      \draw[ring,planeone] (gpu1)--(gpu3);
      \draw[ring,planeone] (gpu3) to[bend left=45] (gpu0);
      \draw[ring,planezero] (-1.1,-.85)--(-.8,-.85);
      \node[anchor=west,text=planezero,font=\sffamily\fontsize{7}{7.6}\selectfont]
        at (-.75,-.85) {Plane 0};
      \draw[ring,planeone] (.25,-.85)--(.55,-.85);
      \node[anchor=west,text=planeone,font=\sffamily\fontsize{7}{7.6}\selectfont]
        at (.6,-.85) {Plane 1};
    \end{scope}
    \draw[ctl] (controller.south)--node[edgecaption,left] {configure}
      (controller.south |- topology.north);
    \draw[payload] (library.south)--node[edgecaption,right,xshift=2pt] {logical exchanges} (transport.north);
    \coordinate (topology-io) at (topology.east |- transport.center);
    \draw[payload] ($(transport.west)+(0,.18)$)
      --node[edgecaption,above] {transfers} ($(topology-io)+(0,.18)$);
    \draw[fault] ($(topology-io)+(0,-.18)$)
      --node[edgecaption,below,yshift=-2pt,align=center,text=red!65!black] {link / transfer\\error}
      ($(transport.west)+(0,-.18)$);
    \end{scope}

    \coordinate (panel-top) at ($(library.north)+(0,.20cm)$);
    \coordinate (panel-bottom) at (transport.south);
    \coordinate (panel-middle) at ($(panel-top)!.5!(panel-bottom)$);

    \begin{scope}[xshift=.8cm,
      edgecaption/.append style={font=\sffamily\fontsize{8}{8.6}\selectfont}]
    \node[decision,anchor=north] (connected) at (12,0 |- panel-top) {Group\\connected?};
    \node[decision] (spread) at (12,0 |- panel-middle) {AR, AG, RS\\and $\Delta h_c>\tau$?};
    \node[decision,anchor=south] (candidate) at (12,0 |- panel-bottom)
      {Available plane\\$\Delta h'_c\leq\tau$?};
    \node[action,draw=planeone,anchor=north] (reconfigure) at (15.35,0 |- panel-top)
      {Reconfigure OCS\\rebuild flow paths};
    \node[action] (keep) at (15.35,0 |- panel-middle) {Keep schedule,\\MRC reroutes};
    \node[action,draw=planezero,anchor=south] (remap) at (15.35,0 |- panel-bottom)
      {Remap channel\\at a boundary};
    \draw[branch] (connected.east)--node[edgecaption,above] {No} (reconfigure.west);
    \draw[branch] (connected.south)--node[edgecaption,left] {Yes} (spread.north);
    \draw[branch] (spread.east)--node[edgecaption,above] {No} (keep.west);
    \draw[branch] (spread.south)--node[edgecaption,left] {Yes} (candidate.north);
    \draw[branch] (candidate.east)--node[edgecaption,above] {Yes} (remap.west);
    \draw[branch] (candidate.north east)-|node[edgecaption,pos=.75,right] {No} (13.65,0 |- keep.center);
    \end{scope}
    \begin{scope}
      \clip (18.1,0 |- panel-bottom) rectangle (22.2,0 |- panel-top);
      \path let \p1=(panel-top), \p2=(panel-bottom), \n1={\y1-\y2} in
        node[anchor=north,inner sep=0] (testbed) at (20.15,0 |- panel-top)
          {\includegraphics[height=\n1]{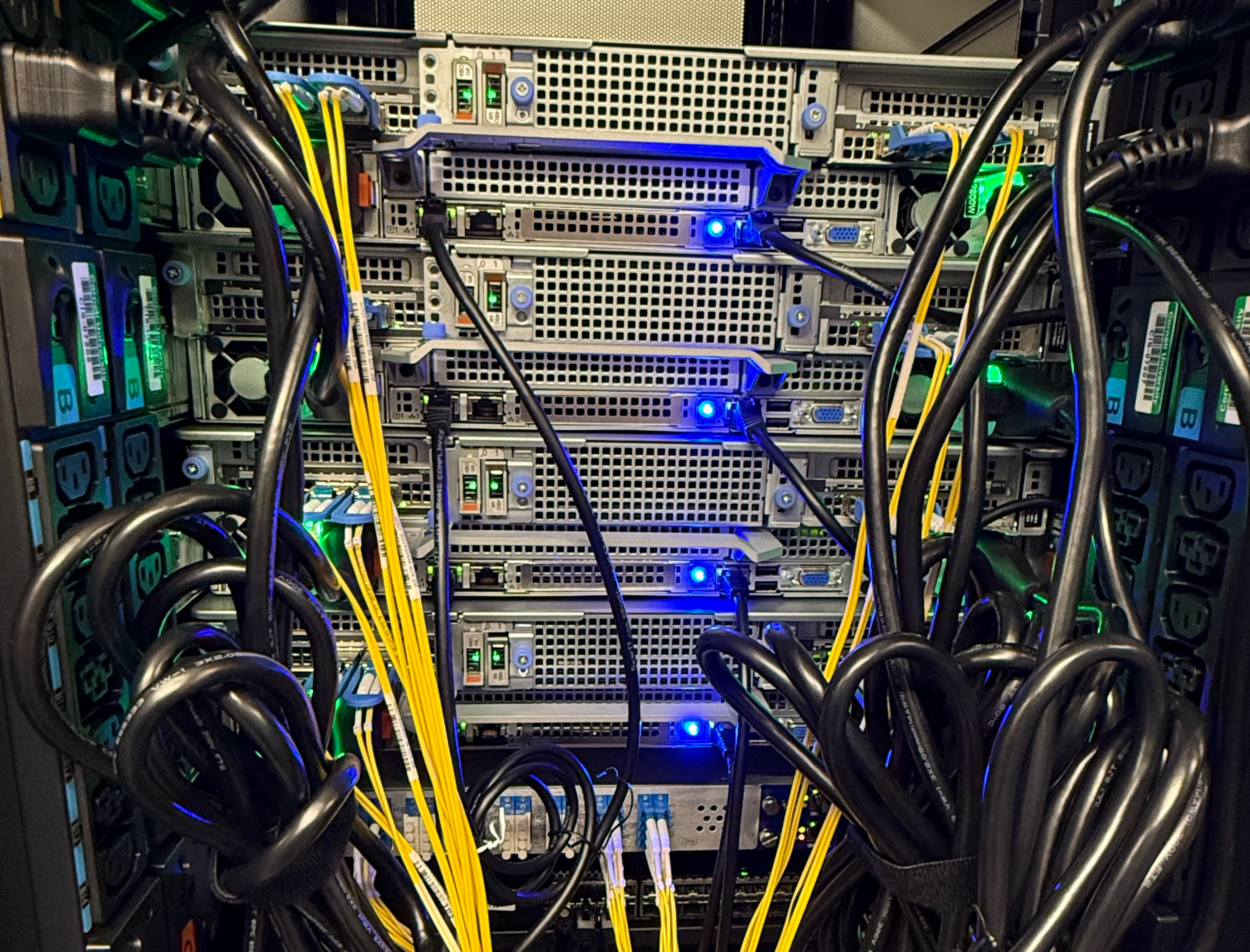}};
      \begin{scope}[shift={(testbed.south west)},
        x={($(testbed.south east)-(testbed.south west)$)},
        y={($(testbed.north west)-(testbed.south west)$)},
        annotation/.style={font=\sffamily\bfseries\fontsize{8}{8.6}\selectfont,
                           text=white,inner sep=1pt}]
        \draw[planezero,line width=1pt] (.12,.205) rectangle (.88,.97);
        \node[annotation,fill=planezero,anchor=north west] at (.12,.97)
          {GPU servers};
        \draw[planeone,line width=1pt] (.16,.04) rectangle (.80,.18);
        \node[annotation,fill=planeone,anchor=south west] at (.16,.04)
          {OCS};
      \end{scope}
    \end{scope}
    \foreach \x/\panel in {5.5/a,14.45/b,20.15/c}
      \node[anchor=north,inner sep=0pt,yshift=-.2cm,
            font=\fontsize{11.5}{12.5}\selectfont]
        at (\x,0 |- panel-bottom) {(\panel)};
    \path[use as bounding box] (0,0 |- panel-top) -- (22.2,0 |- panel-top);
  \end{tikzpicture}}%
  \vspace{-1mm}
  \caption{(a) \sysname architecture. (b) Failure-handling workflow. AR, AG, RS: \allreduce, \allgather, \reducescatter. $\Delta h_c$ is the hop-count spread across flows in channel $c$, and $\tau$ is the threshold. (c) Testbed setup. Four NVIDIA L40 GPUs are connected via a Polatis OCS.}
  \label{fig:sidecar-control}
  \label{fig:testbed}

\end{figure}

\myparab{Collective schedules and transport.}
Within each communication group, the collective traffic is split into \textbf{channels}. Each channel defines a collective schedule specifying the source, destination, data, and timing of each transfer. \sysname maps channels to the planes allocated to their communication group and tailors each channel's schedule to the physical topology. For ring-based collectives, \sysname matches the logical ring ordering to a physical ring permutation so that each scheduled exchange takes one direct hop, as a ring collective produces the same result for any ring order. Separate channels allow processing elements (\eg the streaming multiprocessors of a GPU) to execute transfers in parallel. The collective library issues logical sender--receiver exchanges to the transport layer. A congestion-aware multipath transport (\eg MRC) selects low-hop paths across the group's allocated planes and handles intermediate-node forwarding for traffic such as \alltoall with high fan-in and fan-out.

\myparab{Sidecars and controller.}
A host-side sidecar per accelerator rank collects communication-group information and supplies channel mappings and topology-aware communication orderings to the collective library. A central controller tracks network status, coordinates plane allocations and OCS reconfiguration, and sends topology updates to the sidecars (Fig.~\ref{fig:sidecar-control}(a)). Sidecars apply these updates to collective schedules and transport path sets.

\myparab{Reliable communication within one communication group.}
After a link or plane failure, the transport can preserve a flow's endpoints and use alternative low-hop paths in the group's remaining connectivity, following MRC. For a connected \alltoall group, \sysname keeps the schedule and relies on the transport for rerouting. For each \allreduce channel $c$, let $h_f$ be the shortest available path length for flow $f$, and define the hop-count spread $\Delta h_c=\max_{f\in c}h_f-\min_{f\in c}h_f$. If $\Delta h_c>\tau$ (\eg $\tau=2$ hops), \sysname looks for available planes in the group's allocation whose topology-matched schedule has spread $\Delta h'_c\leq\tau$. If they exist, \sysname remaps the channel at a collective boundary and rebalances the data partition across channels for load balancing. Otherwise, it keeps the schedule and uses transport rerouting (Fig.~\ref{fig:sidecar-control}(b)). The threshold $\tau$ can be tuned to the size of the topology.

\myparab{Reliable communication across communication groups.}
If the affected communication group is disconnected and cannot route all required flows, the controller first reconfigures the OCS to reallocate planes from other groups. A possible donor policy selects the group with the most healthy planes. After reconfiguration, the controller updates plane allocations, and the transport rebuilds the corresponding path sets to recover connectivity (Fig.~\ref{fig:sidecar-control}(b)).

\vspace{-3pt}
\section{Evaluation}
\vspace{-2pt}

\myparab{Setup and methodology.}
We use four hosts, each with one NVIDIA L40 GPU and four 100~Gb/s remote direct memory access (RDMA) ports (two per ConnectX-6 Dx NIC, equipped with 4$\times$25~Gb/s non-return-to-zero (NRZ) 1310~nm transceivers) connected through a 32$\times$32 Polatis 6000n OCS (Fig.~\ref{fig:testbed}(c)).
Our prototype uses MSCCL++ for collectives and RDMA over Converged Ethernet v2 (RoCEv2) for transport. We use a software-emulated multipath transport with GPU store-and-forward in both \sysname and the MRC baseline because our NICs lack the required forwarding capability.
We deploy \texttt{Qwen1.5-MoE-A2.7B-Chat}, which has 14.3B parameters, 24 layers, and 60 experts per layer with top-4 routing, using WikiText-2 tokens.

\begin{figure}[t]
  \centering
  \begin{minipage}[b]{0.245\linewidth}
    \centering
    \includegraphics[width=\linewidth]{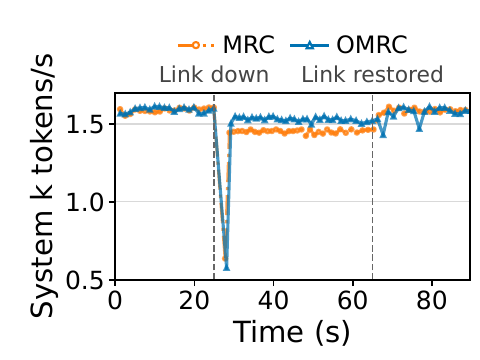}\\[-4pt]
    \makebox[\linewidth][l]{\hspace*{0.24\linewidth}%
      \makebox[0.74\linewidth]{\small (a) Training, $\tau=0$}}
  \end{minipage}\hfill
  \begin{minipage}[b]{0.245\linewidth}
    \centering
    \includegraphics[width=\linewidth]{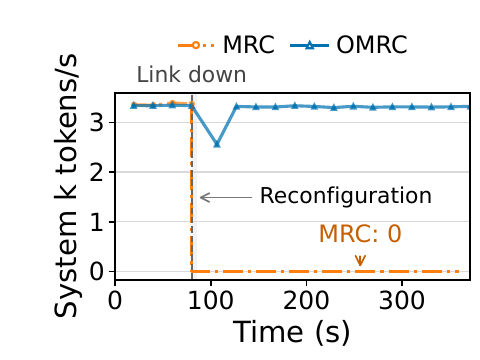}\\[-4pt]
    \makebox[\linewidth][l]{\hspace*{0.24\linewidth}%
      \makebox[0.74\linewidth]{\small (b) Inference, $\tau=0$}}
  \end{minipage}\hfill
  \begin{minipage}[b]{0.245\linewidth}
    \centering
    \includegraphics[width=\linewidth]{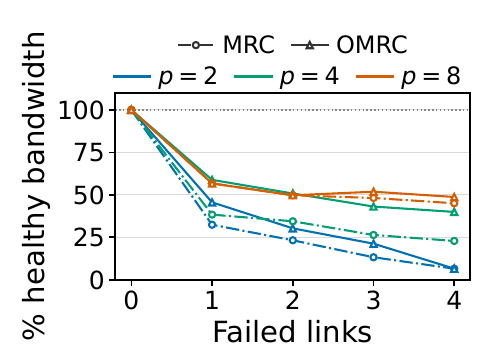}\\[-4pt]
    \makebox[\linewidth][l]{\hspace*{0.24\linewidth}%
      \makebox[0.74\linewidth]{\small (c) $n=16$, $\tau=2$}}
  \end{minipage}\hfill
  \begin{minipage}[b]{0.245\linewidth}
    \centering
    \includegraphics[width=\linewidth]{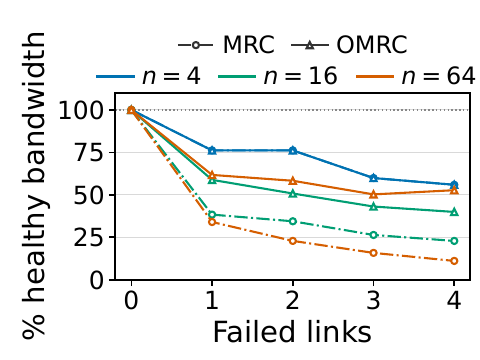}\\[-4pt]
    \makebox[\linewidth][l]{\hspace*{0.24\linewidth}%
      \makebox[0.74\linewidth]{\small (d) $p=4$, $\tau=2$}}
  \end{minipage}
  \vspace{-2mm}
  \caption{(a) Training with expert and tensor parallelism (EP/TP). (b) Disaggregated EP/TP inference. (c,d) 
  Simulated 64~MB \allreduce failure scaling with (c) 16 ranks and (d) 4 planes. The results are the means from four sampling runs.}
  \label{fig:workload-evaluation}
  \label{fig:allreduce-failure-scaling}
  \vspace{-2mm}
\end{figure}

\myparab{Experiment 1: temporary link failure during training.}
We use overlapping four-rank tensor and expert parallelism (TP4/EP4), with a global batch of $8\times256$ tokens.
TP outputs and backward input gradients use \allreduce. Expert dispatch/combine use \alltoall. Four channels in the TP/EP group span two planes, utilizing bidirectional bandwidth on each link. We cut plane~0's link $0$--$1$ after 25~s by programming the switch and restore it 40~s later (Fig.~\ref{fig:workload-evaluation}(a)). When the link cut is detected, MRC reroutes the impacted flow, whereas \sysname remaps the impacted channel to the remaining planes.
Remapping reduces the mean hop count from 1.125 to 1.0 and increases the median 8~MiB \allreduce algorithmic bandwidth from 4.00 to 4.97~GB/s (24\%), while \alltoall bandwidth differs by less than 4\% between the two policies.
The steady degraded-phase throughput is 1.45k tokens/s for MRC and 1.53k for \sysname, a 5.3\% gain. \sysname retains 95.7\% of its healthy throughput.

\myparab{Experiment 2: permanent link failure during disaggregated inference.}
We place prefill on ranks $[0,1]$ and decode on $[2,3]$. Each stage runs TP2/EP2 with one plane. Each KV-cache transfer group, $[0,2]$ and $[1,3]$, initially has three planes. The inference batch contains 32 prompts of 2,048 tokens and 64 generated tokens per request. Each batch transfers 12~GiB of KV state. We disable the sole TP/EP plane after 80~s (Fig.~\ref{fig:workload-evaluation}(b)). The job using MRC stalls. \sysname reconfigures the OCS to reallocate one plane from the KV pairs to restore the TP/EP group. Median throughput decreases slightly from 3.335k to 3.311k tokens/s (0.7\%), and the job keeps running.

\myparab{Simulation.}
We extend ASTRA-sim's congestion-aware analytical network backend~\cite{Won_2023} to model direct-connect multi-plane optical DCs.
We implement a multipath transport that distributes sends across equal-cost paths on planes allocated to the collective, and an \sysname controller that manages logical ring orderings, channel remapping, and physical reconfiguration of the network in response to failures.
In Fig.~\ref{fig:allreduce-failure-scaling}(c,d), we perform a sweep across GPU counts and plane configurations with compounding failed links. Failures are determined by randomly selecting a link and disabling it. Each configuration is run four times, and the lines show the mean.
Figure~\ref{fig:allreduce-failure-scaling}(d) shows that \sysname provides a larger improvement over MRC for larger communication groups. With four planes and four failed links, \sysname improves \allreduce bandwidth over MRC by $1.74\times$ at 16 ranks and $4.74\times$ at 64 ranks.


\vspace{-3pt}
\section{Conclusion}
\vspace{-2pt}
\sysname extends MRC and combines topology-aware collective remapping with OCS plane reallocation to recover from failures in direct-connect optical interconnects. On our testbed, it retains 95.7\% of healthy training throughput and restores 99.3\% of pre-failure inference throughput where MRC stalls. In simulation, \sysname achieves up to $4.74\times$ MRC's \allreduce bandwidth at 64 ranks under compounding link failures.

\vspace{-4pt}
\bibliographystyle{opticajnl-compact}
\bibliography{references}

@misc{mrc,
  author = {Joao Araujo and Alex Chow and Mark Handley and Ryder Lewis and
            Christoph Paasch and Jitendra Padhye and Michael Papamichael and
            Greg Steinbrecher and Amin Tootoonchian and Lihua Yuan and
            S. Anantharamu and Abhishek Dosi and Mohit Garg and Mahdieh Ghazi and
            Torsten Hoefler and Deepal Jayasinghe and Jithin Jose and
            Abdul Kabbani and Guohan Lu and Yang Wang and K. Doddapaneni and
            Murali Garimella and Vipin Jain and Yanfang Le and H. Nagulapalli and
            S. Narayanan and Rong Pan and Rathina Sabesan and Raghava Sivaramu and
            Rip Sohan and Eric Davis and Dragos Dumitrescu and Mohan Kalkunte and
            Bhaswar Mitra and Guglielmo Morandin and Adrian Popa and Costin Raiciu and
            Eric Spada and John Spillane and Niranjan Vaidya and Aviv Barnea and
            Idan Burstein and Elazar Cohen and Yamin Friedman and Noam Katz and
            Masoud Moshref and Yuval Shpigelman and Shahaf Shuler and Shy Shyman and
            Sayantan Sur},
  title = {Resilient {AI} Supercomputer Networking using {MRC} and {SRv6}},
  year = {2026},
  eprint = {2605.04333},
  archivePrefix = {arXiv},
  primaryClass = {cs.NI},
  doi = {10.48550/arXiv.2605.04333},
  howpublished = {arXiv:2605.04333}
}

@inproceedings{actina,
  title={{ACTINA}: adapting circuit-switching techniques for {AI} networking architectures},
  author={Wu, Zhenguo and Klenk, Benjamin and Dennison, Larry and Bergman, Keren},
  booktitle={Proceedings of the International Conference for High Performance Computing, Networking, Storage and Analysis},
  pages={1211--1222},
  year={2025}
}

@inproceedings{opus,
  author = {Eric Ding and Barry Lyu and Bhaskar Kataria and Rachee Singh},
  title = {{Opus}: Photonic Rail-Optimized Fabric in {ML} Datacenters},
  booktitle = {Proceedings of the ACM SIGCOMM 2026 Conference},
  series = {SIGCOMM '26},
  year = {2026},
  publisher = {ACM},
  pages = {922--939},
  doi = {10.1145/3789240.3829187},
}

@inproceedings{topoopt,
  title={{TopoOpt}: Co-optimizing network topology and parallelization strategy for distributed training jobs},
  author={Wang, Weiyang and Khazraee, Moein and Zhong, Zhizhen and Ghobadi, Manya and Jia, Zhihao and Mudigere, Dheevatsa and Zhang, Ying and Kewitsch, Anthony},
  booktitle={20th USENIX Symposium on Networked Systems Design and Implementation (NSDI 23)},
  pages={739--767},
  year={2023}
}

@inproceedings{Won_2023,
  title={{ASTRA-sim2.0}: Modeling hierarchical networks and disaggregated systems for large-model training at scale},
  author={Won, William and Heo, Taekyung and Rashidi, Saeed and Sridharan, Srinivas and Srinivasan, Sudarshan and Krishna, Tushar},
  booktitle={2023 IEEE International Symposium on Performance Analysis of Systems and Software (ISPASS)},
  pages={283--294},
  year={2023},
  organization={IEEE}
}

\end{document}